\documentclass[conference]{IEEEtran}
\IEEEoverridecommandlockouts

\usepackage{cite}
\usepackage{amsmath,amssymb,amsfonts}
\usepackage{algorithmic}
\usepackage{graphicx}
\usepackage{booktabs}
\usepackage{textcomp}
\usepackage{xcolor,balance}
\usepackage{graphicx}
\usepackage{subcaption} 
\usepackage{tabularx}
\usepackage{booktabs} 
\usepackage{multirow}
\usepackage{pifont}  
\usepackage{balance}
\def\BibTeX{{\rm B\kern-.05em{\sc i\kern-.025em b}\kern-.08em
    T\kern-.1667em\lower.7ex\hbox{E}\kern-.125emX}}
    
\begin{document}

\title{Air-Sea Surface Modeling and Operating Link Range Evaluation for AUV-to-UAV Optical Wireless Communication Links \\

\thanks{This work was supported in part by the Séjours Scientifiques de Haut Niveau (SSHN) 2025 Research Grant from the Embassy of France in Nigeria, the Invited Lecturer Fellowship of École Centrale Méditerranée (ECM), and Alex Ekwueme Federal University Ndufu-Alike (AE-FUNAI).}
}

\makeatletter
\newcommand{\linebreakand}{%
\end{@IEEEauthorhalign}
\hfill\mbox{}\par
\mbox{}\hfill\begin{@IEEEauthorhalign}
}
\makeatother

\author{\IEEEauthorblockN{Ikenna Chinazaekpere Ijeh}
	\IEEEauthorblockA{\textit{Department of Electrical/Electronic Engineering} \\
		\textit{Alex Ekwueme Federal University, Ndufu-Alike}\\
		Ebonyi State, Nigeria \\
		ikenna.ijeh@funai.edu.ng}
	\and
	\IEEEauthorblockN{Mohammad Ali Khalighi}
	\IEEEauthorblockA{\textit{Aix Marseille University, CNRS} \\
		\textit{Centrale Méditerranée, Institut Fresnel}\\
		Marseille, France \\
		ali.khalighi@fresnel.fr}
	\and
	\IEEEauthorblockN{Wasiu O. Popoola}
	\IEEEauthorblockA{\textit{School of Engineering} \\
		\textit{University of Edinburgh}\\
		Edinburgh, United Kingdom\\
		w.popoola@ed.ac.uk}
}

\maketitle

\begin{abstract}
Air-sea surface interactions play a critical role in underwater-to-air optical wireless communication (OWC) links, particularly in vertical autonomous underwater vehicle (AUV) to unmanned aerial vehicle (UAV) scenarios, where the stochastic nature of the sea surface introduces optical distortions that impair link reliability. This work investigates the impact of air-sea surface roughness on AUV-to-UAV OWC systems using two experimentally validated models: the classical Cox-Munk and the Elfouhaily-Chapron-Katsaros-Vandemark (ECKV).
A tractable analytical representation of the ECKV model is derived and validated against measured sea-state data. Using both analytical and Monte Carlo approaches, the link ergodic capacity is evaluated with particular emphasis on operating range, pointing errors, receiver field-of-view, and solar noise level, providing practical system design insights. 
\end{abstract}


\section{Introduction}\label{Sec-Intro}

The proliferation of autonomous marine and aerial platforms has created a growing demand for high-capacity, low-latency communication between underwater and above-water systems \cite{angara2024influence}. Applications such as coordinated autonomous underwater vehicle (AUV)-unmanned aerial vehicle (UAV) missions, real-time oceanographic sensing, environmental monitoring, and offshore infrastructure inspection, require timely transmission of large data volumes across the water-air interface for centralized processing and control \cite{angara2024influence,ijeh2023optical}.

Conventional communication technologies, acoustic underwater and radio-frequency (RF) in air, are inherently unsuitable for direct underwater-to-air links because they experience extremely high attenuation when propagating through the opposite medium. \textcolor{black}{This limitation is more pronounced at moderate water-to-air ranges, where RF signals are strongly absorbed in water and acoustic waves suffer from impedance mismatch at the interface. This motivates optical wireless communication (OWC) as a high-speed, low-latency alternative capable of enabling cross-medium links over tens of meters in both water and air using the visible spectrum \cite{Kaushal-Access-2016}.} Nevertheless, direct underwater-to-air OWC remains relatively underexplored, mainly due to the complexity of air-sea interface, where refractive-index discontinuities, dynamic surface roughness, and wave-induced refraction can induce beam distortion, pointing errors, and severe power fluctuations \cite{angara2024influence,Ijeh-JOCN-2022,angara2024performance}, making vertical air-water optical channels particularly challenging.

A widely used model for the air-sea interface is the Cox-Munk (CM) model, which empirically describes sea surface slope statistics as a function of wind speed, providing a first-order representation of surface roughness \cite{Ijeh-JOE-2021, ijeh2022ergodic}. While attractive for its simplicity, the CM model does not explicitly capture spatial and temporal correlations in wave dynamics, limiting its accuracy under moderate-to-high wind conditions \cite{angara2024influence, guerin2023revisiting}. 
More models, such as the Elfouhaily-Chapron-Katsaros-Vandemark (ECKV) spectrum, incorporate detailed descriptions of wind-driven wave fields, including multi-scale interactions, directionality, and spatio-temporal correlations \cite{Elfouhaily-JGeoR-1997}. However, the complexity of the ECKV model limits its direct use in analytical link design and performance evaluation.

\textcolor{black}{To address this limitation, we introduce a tractable analytical model,
based on the validated ECKV formulation to match real stereo-image measurements under diverse wind and sea conditions \cite{angara2024influence}. The model is selected based on best-fit performance among existing distributions and integrated into a vertical AUV–UAV OWC framework to characterize air–sea interface effects, while preserving realistic dynamics and enabling analytical performance evaluation. Using this, we quantify the joint impact of surface roughness, pointing errors, and receiver field-of-view on ergodic capacity, highlighting range-dependent regimes and key design trade-offs.}



Subsequently, Section~\ref{Sec-Sys-Chn} describes the underwater-to-air system model and assumptions. Section~\ref{Sec-A2S-Mdl} presents the proposed air-sea surface model, while Section~\ref{Sec-Metric} develops the link ergodic capacity analysis. Numerical results are discussed in Section~V, and conclusions are drawn in Section~\ref{Sec-Concl}.

\begin{figure}[t]
	\centering
	\includegraphics[scale=0.4]{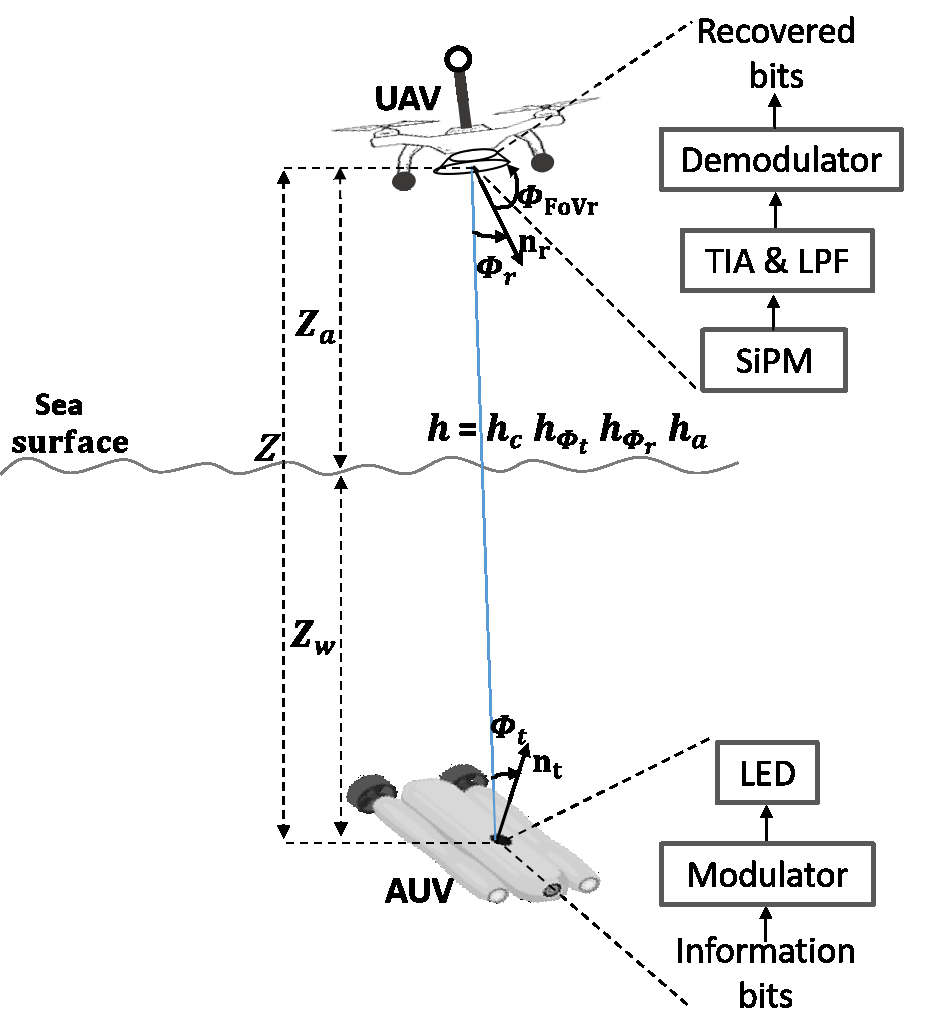}
	\caption{\small Considered direct water-to-air OWC scenario.}
	\label{fig:W2Aillust}
\end{figure}

\begin{figure*}[t]
	\centering
	\begin{subfigure}[t]{0.32\textwidth}
		\centering
		\includegraphics[scale=0.44]{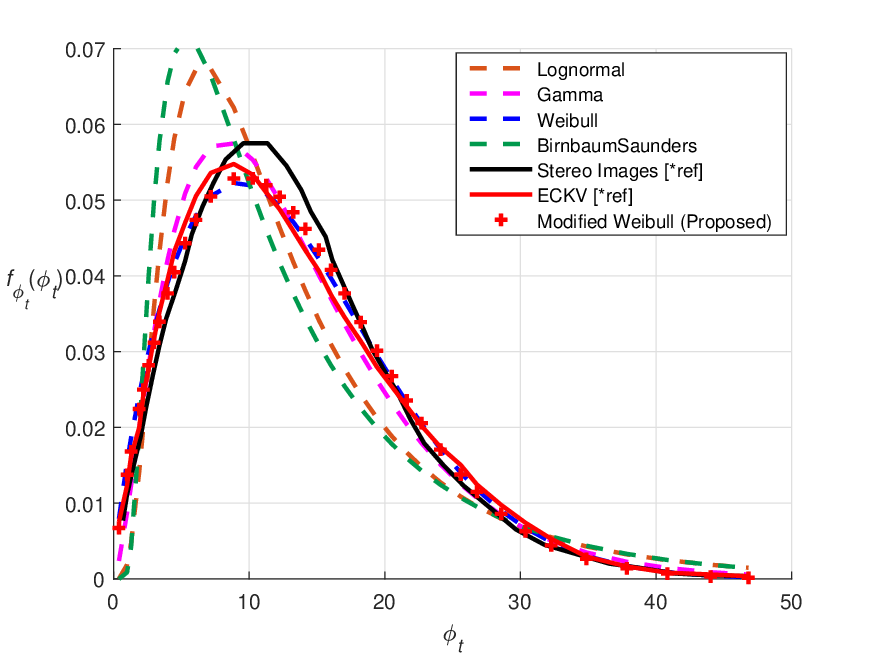}
		\caption{\small Black Sea; $U = 6.1$\,\text{m/s}}
	\end{subfigure}
	\begin{subfigure}[t]{0.32\textwidth}
		\centering
		\includegraphics[scale=0.44]{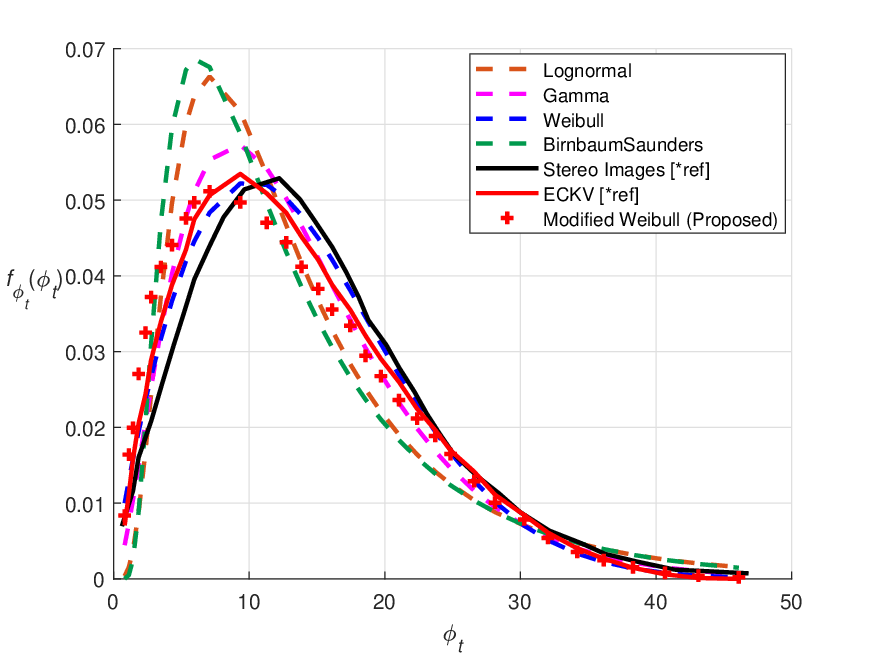}
		\caption{\small Black Sea; $U = 8.7$\,\text{m/s}}
	\end{subfigure}
	\begin{subfigure}[t]{0.32\textwidth}
		\centering
		\includegraphics[scale=0.44]{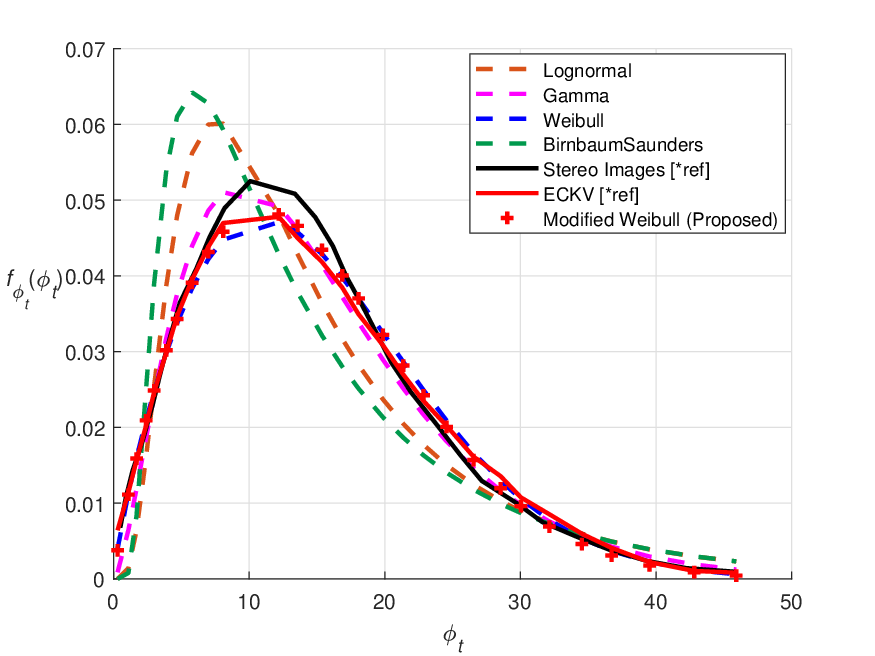}
		\caption{\small Black Sea; $U = 12.9$\,\text{m/s}}
	\end{subfigure}
	\vspace{0.4cm}
	
	\begin{subfigure}[t]{0.32\textwidth}
		\centering
		\includegraphics[scale=0.44]{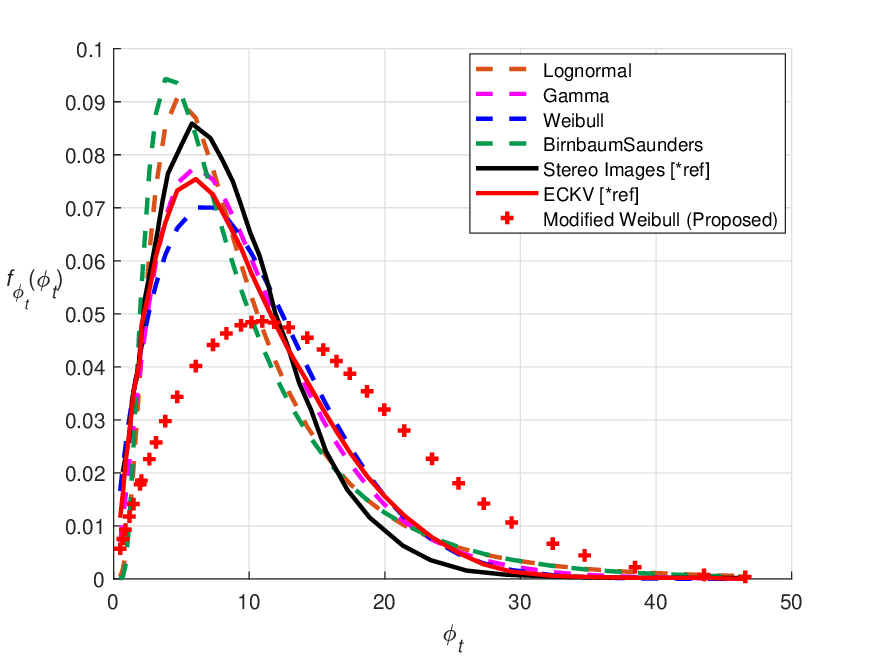}
		\caption{\small Adriatic Sea; $U = 12.9$\,\text{m/s}}
	\end{subfigure}
	\begin{subfigure}[t]{0.32\textwidth}
		\centering
		\includegraphics[scale=0.44]{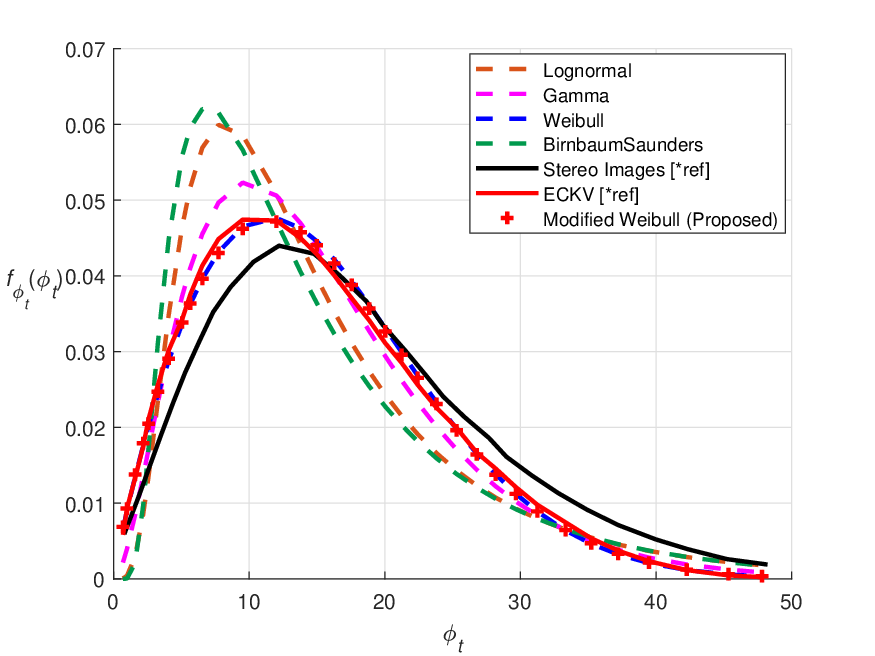}
		\caption{\small Black Sea; $U = 15.2$\,\text{m/s}}
	\end{subfigure}
	\begin{subfigure}[t]{0.32\textwidth}
		\centering
		\includegraphics[scale=0.44]{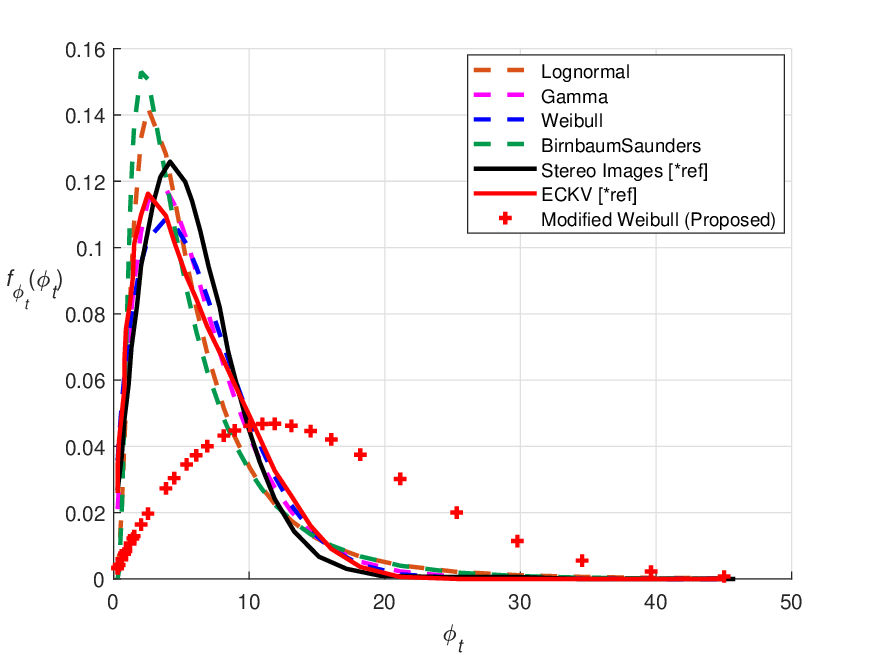}
		\caption{\small Yellow Sea; $U = 16$\,\text{m/s}}
	\end{subfigure}
	
	\caption{\small PDF of $\phi_t$, $f_{\phi_{t}}(\phi_{t})$ for various seas and wind speeds (\textit{[*ref]: Data for both stereo images and ECKV were reproduced from \cite{angara2024influence}}).}
	\label{fig:pdf_phi_windspeeds}
\end{figure*}

\section{System Model and Channel Assumptions}\label{Sec-Sys-Chn}

\subsection{AUV-UAV Optical Link Geometry}
Figure~\ref{fig:W2Aillust} depicts the vertical underwater-to-air OWC link under study. An AUV transmits optical data upward through the dynamic air-sea interface to a hovering UAV. The total link range is denoted by $Z = Z_w + Z_a$, where $Z_w$ and $Z_a$ denote the underwater depth and atmospheric height, respectively.
Obviously, the relative positioning of the AUV and UAV directly impacts system performance. 
At low altitudes, rotor-induced airflow further disturb the sea surface, increasing roughness, beam distortion, and glint variability. Meanwhile, the AUV depth determines the extent of underwater attenuation, scattering, and beam spreading prior to surface emergence.

The transmitter (Tx) and receiver (Rx) optical axes are assumed to be nominally vertical, with unit normal vectors $\mathbf{n}_{\mathrm{t}}$ and $\mathbf{n}_{\mathrm{r}}$, respectively. 
We assume that the Tx irradiance angle $\phi_t \in [0,\pi/2]$ is primarily governed by the instantaneous sea-surface slope variations. Also, for simplicity, we ignore AUV instability and underwater current effects, corresponding to stabilized hovering operation. Accordingly, $\phi_t$ follows the sea-surface slope statistics.
At the Rx, misalignment arises from UAV motion and wind-induced vibrations. For simplicity, the Rx incidence angle $\phi_r \in [0,\pi/2]$ is modeled as a zero-mean Gaussian random variable with variance $\sigma_{\phi_r}^2$, truncated by Rx field-of-view (FoV), $\phi_\text{FoV}$. Note, perfect beam alignment corresponds to $\phi_t = \phi_r = 0$.

\subsection{Signal Propagation and Channel Model}
Intensity modulation with direct detection (IM/DD) using on-off keying (OOK) is employed. At the Tx (AUV), the optical source, a light-emitting diode (LED), is modeled as a generalized Lambertian emitter of order $m$. 
At the Rx  (UAV), a highly-sensitive silicon photomultiplier (SiPM) is used to convert the  optical power into a photocurrent, which is amplified by a trans-impedance amplifier (TIA), and low-pass filtered (LPF), before signal demodulation, see Fig.~\ref{fig:W2Aillust}.
The overall channel attenuation $h$ is the product of deterministic path loss $h_c$ and angular-dependent factors \cite{Ijeh-JOE-2021}:

{\small
\begin{equation}\label{Eq:h-complete}
	\begin{split}
		h  &=  \underbrace{\frac{{m}+1}{2\pi}\frac{A_\text{PD}\, T_\text{s}\, g}{Z^2}\, e^{-{(Z\,K}_\text{eff})}}_{\displaystyle h_c} \underbrace{\cos^{m}(\phi_\text{t})}_{\displaystyle h_{\phi_{\text{t}}}} \underbrace{\cos(\phi_\text{r})}_{\displaystyle h_{\phi_{\text{r}}}} \underbrace{\Pi\Big(\frac{\phi_\text{r}}{\phi_\text{FoV}}\Big)}_{\displaystyle h_a}.
	\end{split}
\end{equation}
}
Here, $h_c$, accounts for distance-dependent attenuation, where $A_{\mathrm{PD}}$ is the photodetector \textcolor{black}{(PD)} active area, $T_s$ is the optical filter transmittance, $g=\frac{n_{rf}^2}{\sin^2(\phi_{\mathrm{FoV}})}$ is the concentrator gain with refractive index $(n_{rf})$. Also, $K_\text{eff}=(K_w Z_w + K_a Z_a)/Z$, denotes the effective water-air attenuation coefficient, with $K_w$ and $K_a$ the diffuse attenuation coefficients of water and air, respectively. 
The terms $h_{\phi_{\text{t}}}$ and $h_{\phi_{\text{r}}}$ account for angular misalignment, while $h_a$ enforces the Rx FoV constraint, i.e., $\Pi(\phi_{r}/\phi_\text{FoV})=1$ if $\phi_{r}\leq \phi_\text{FoV}$, and 0 otherwise. 

\subsection{Air-Sea Surface Modeling}\label{Sec-A2S-Mdl}
In air-sea optical wireless communication links, wind-driven surface roughness introduces stochastic pointing errors that can dominate link performance. 
To model this effect, two widely used approaches are the CM and ECKV models that we briefly introduce in the following.

\paragraph{CM Model}
Based on extensive optical measurements of Sun glitter over the ocean, the CM approach models
surface slope components as random variables with
variances that depend linearly on the wind speed, $U$, within the operating range of $1 - 14$~m/s.
The probability density function (PDF) of $\phi_t$ for the CM model can be expressed as \cite{Zhang-LCOMM-2015}:
\begin{equation}\label{eq:phitpdf_CM}
	f_{\phi_\text{t}}(\phi_\text{t})=\frac{\tan\left(\phi_\text{t}\right)\sec^2\left(\phi_\text{t}\right)}{2\pi\sigma_{U}^2}\exp\left(\frac{-\tan^2\left(\phi_\text{t}\right)}{2\sigma_{U}^2}\right),
\end{equation}
where $\sigma_U^2 = 0.003 + 0.00512~U$.
Although attractive for its simplicity, the CM model lacks explicit representation of spatial and temporal surface correlations and multi-scale wave interactions, thereby restricting its accuracy.
\paragraph{ECKV Model}
In contrast to the CM model, the more recent ECKV spectrum provides a unified, physically consistent two-dimensional sea-surface model that captures spatial correlations across both gravity and capillary wave scales, offering more realistic approximations \cite{Elfouhaily-JGeoR-1997}. It characterizes sea-surface elevations and slope statistics with respect to the wind speed through the directional wave spectrum given as \cite{angara2024influence}:
\begin{equation}
	\psi(k,\phi)=\frac{1}{k} S(k)\,D(k,\phi),
\end{equation}
where $S(k)$ is the omnidirectional elevation-variance spectrum, $D(k,\phi)$ is the angular spreading function, and $k$ denotes the spatial wavenumber and $\phi$ is the angle relative to the direction of wave propagation and the wind. 

In \cite{angara2024influence}, the ECKV model was investigated using real stereo-image measurements acquired under diverse wind and sea conditions, for datasets from the Black Sea \cite{guimaraes2018sea}, Adriatic Sea \cite{benetazzo2018stereo}, and Yellow Sea \cite{zhang1999flexible}, to study its capability to accurately represent realistic sea-surface roughness. The realized results for empirical PDF $f_{\phi_{t}}(\phi_{t})$, shown in Fig.\,\ref{fig:pdf_phi_windspeeds}, demonstrates strong agreement between distributions from the stereo measurements and that predicted by the ECKV model (shown in black and red, respectively).

\begin{table}[t]
	\centering
	\caption{\small MSE comparison of ECKV data with considered PDFs at various wind speeds $U$ [\textit{MSE values are in $\times 10^{-6}$}]}
	\resizebox{\columnwidth}{!}{%
		\begin{tabular}{|c|c|c|c|c|c|}
			\hline
			\textbf{Fig.}
			& \textbf{\textit{U} (m/s)} 
			& \textbf{Lognormal} 
			& \textbf{Gamma} 
			& \textbf{Weibull}
			& \textbf{B-S} \\ \hline 	
			\ref{fig:pdf_phi_windspeeds}a & 6.1  & 47.27  & 4.25  & 3.75  & 109.21 \\
			\ref{fig:pdf_phi_windspeeds}b & 8.7  & 58.76  & 7.60  & 2.42 & 95.29  \\
			\ref{fig:pdf_phi_windspeeds}c & 12.9 & 61.54  & 7.67  & 0.94  & 141.92 \\
			\ref{fig:pdf_phi_windspeeds}d & 12.9 & 72.01  & 4.01 & 9.83  & 139.28 \\
			\ref{fig:pdf_phi_windspeeds}e & 15.2 & 54.91  & 7.09  & 0.702  & 96.46  \\
			\ref{fig:pdf_phi_windspeeds}f & 16.0 & 233.91 & 34.73 & 37.27 & 433.94 \\ \hline
		\end{tabular}
	}
	\label{tab:mse_windspeed}
\end{table}


%

\section{Proposed Air-Sea Surface Model}\label{Sec-A2S-Mdl}

In an attempt to obtain an analytically tractable expression for $f_{\phi_{t}}(\phi_{t})$ based on the ECKV model, first, we performed curve fitting of the ECKV data, using parametric distributions commonly employed in wireless fading analysis, including Lognormal, Gaussian, Exponential, Gamma, Weibull, and Birnbaum-Saunders (B-S). We have presented in Fig.~\ref{fig:pdf_phi_windspeeds} the most representative fits alongside the empirical ECKV PDF. 
Then, a specific sea region was selected to ensure consistent data availability and reconstruction resolution. Namely, we used  Black Sea measurements for wind speeds of $6.1$, $8.7$, and $15.2$\,\text{m/s} having resolution of $0.1$\,\text{m} (the $U=12.9$\,\text{m/s} dataset was excluded due to its different $0.05$\,\text{m} resolution). 
As illustrated in Figs.~\ref{fig:pdf_phi_windspeeds}(a), (b), and (e), the Weibull distribution shows strong visual agreement with the empirical ECKV data across these wind speeds. This observation is corroborated by the mean squared error (MSE) values presented in Table~\ref{tab:mse_windspeed}, where the Weibull fit attains the lowest MSEs of $3.75\times10^{-6}$, $2.42\times10^{-6}$, and $0.702\times10^{-6}$ for $U=6.1$, $8.7$, and $15.2$~m/s, respectively. Compared with  Lognormal, Gamma, and B-S models, these consistently minimal errors indicate that the Weibull distribution provides the closest statistical approximation to the ECKV data over a wide range of wind conditions.

Focusing now on the Weibull distribution, it has two parameters of shape $k$ and scale $\lambda$. The calculated values of these parameters for the considered wind speeds are ($k_{6.1}=1.77$, $\lambda_{6.1}=15.21$),  ($k_{8.7}=1.84$, $\lambda_{8.7}=15.61$), and ($k_{15.2}=1.84$, $\lambda_{15.2}=17.34$). The next step was to establish a relationship between the wind speed $U$ and the Weibull parameters $k$ and $\lambda$ in order to obtain a generalized model. To this end, we investigated linear and power-law regression models and calculated their parameters using least-squares estimation. 
The resulting relationships are:
\begin{equation}\label{Eq:kU-lU-IE}
	{k_{U} = 1.7454 + 0.0071\,U}, 
	\ \
	{\lambda_{U} = 13.6485 + 0.2406\,U},
\end{equation}
\begin{equation}
	{k_U = 1.6506\,U^{0.0428}}, 
	\quad
	{\lambda_{U} = 11.4724\,U^{0.1499}}. 
\end{equation}
\begin{table}
	\centering
	\caption{\small Comparison between fitted Weibull parameters with linear and power-law regression approximations and associated MAE}
	\renewcommand{\arraystretch}{1.2}
	\begin{tabularx}{\columnwidth}{|c|*{6}{>{\centering\arraybackslash}X|}}
		\hline
		\textbf{\textit{U} (m/s)} 
		& \multicolumn{3}{c|}{\textbf{\textit{k}}}
		& \multicolumn{3}{c|}{$\boldsymbol{\lambda}$} \\ \cline{2-7}
		
		& $k_{\text{actual}}$ & $k_{\text{linear}}$ & $k_{\text{power}}$ & $\lambda_{\text{actual}}$ & $\lambda_{\text{linear}}$ & $\lambda_{\text{power}}$ \\
		\hline
		6.1  & 1.7671 & 1.7887 & 1.7835 & 15.2100 & 15.1160 & 15.0444  \\
		8.7 & 1.8373 & 1.8071 & 1.8108 & 15.6100 & 15.7415 & 15.8668 \\
		15.2 & 1.8446 & 1.8532 & 1.8546 & 17.3429 & 17.3053 & 17.2509  \\
		\hline
		\textbf{MAE}
		& $\text{MAE}_{k}$
		& \textbf{Linear: 0.0209}
		& \textbf{Power: 0.0196} 
		& $\text{MAE}_{\lambda}$
		& \textbf{Linear: 0.0250}
		& \textbf{Power: 0.0402}\\
		\hline
	\end{tabularx}
	\label{tab:mae_weibull}
\end{table}
%
Finally, identifying the most suitable parametric model, the estimated Weibull parameters were quantitatively compared with their reference (actual) values using the mean absolute error (MAE), averaged over the three wind speeds, see Table~\ref{tab:mae_weibull}. As shown, both linear and power-law regressions capture the expected monotonic increase of the shape parameter $k$ and scale parameter $\lambda$ with wind speed, consistent with increased surface roughness under stronger winds. However, their accuracy differs: the power-law model yields a slightly lower MAE for $k$ ($0.0196$ versus $0.0209$), whereas the linear model provides a substantially lower MAE for 
$\lambda$ ($0.0250$ versus $0.0402$).
Since $\lambda$ governs the spread and magnitude of the Weibull distribution and thus has greater impact on the statistical characterization, we adopt the linear model.
In fact, its superior accuracy for 
$\lambda$, combined with its simplicity and interpretability, makes it the more practical choice for modeling wind-speed-dependent Weibull parameters.



Based on these, we derive an analytical expression for the PDF of  $\phi_{t}$ conditioned on wind speed $U$ using a modified Weibull distribution, referred to as the \textcolor{black}{MW} model, given by:
\begin{equation}\label{eq:phitpdf_IE}
f_{\phi_{t}}(\phi_{t})=\frac{k_{U}}{\lambda_{U}}
\Big(\frac{\phi_{t}}{\lambda_{U}}\Big)^{k_{U}-1}
\exp\Big[-\Big(\frac{\phi_{t}}{\lambda_{U}}\Big)^{k_{U}}\Big],
\end{equation}
\textcolor{black}{where $k_{U}$ and $\lambda_{U}$ are given by (\ref{Eq:kU-lU-IE}).}

Under the considered Black Sea wind conditions, including $12.9$~m/s, which was not used in model fitting, the \textcolor{black}{MW} model shows close agreement with the ECKV data (Figs.~\ref{fig:pdf_phi_windspeeds}(a), (b), (c), and (e)), indicating validity beyond calibration points and an effective applicability range of roughly $6-15$~m/s.
In contrast, for the Adriatic and Yellow Sea datasets, the \textcolor{black}{MW}  model does not match the ECKV statistics despite similar wind speeds, because its Weibull scaling parameters were derived under different conditions, characterized by lower $f_{\phi_t}(\phi_t)$ magnitudes and a higher occurrence of large tilt angles at stronger winds (see Fig.~\ref{fig:pdf_phi_windspeeds}(c) vs. (d) and Fig.~\ref{fig:pdf_phi_windspeeds}(e) vs. (f)). 
Indeed, Table~\ref{tab:mse_windspeed} indicates that, for these cases, the Gamma distribution provides the lowest MSE. This highlights the influence of region-specific sea-state and wind-wave dynamics on surface-slope statistics. Nevertheless, the proposed modeling framework remains adaptable, as its parameters can be recalibrated for other environments using the same methodology.

\begin{figure}
	\centering
	\includegraphics[width=\linewidth]{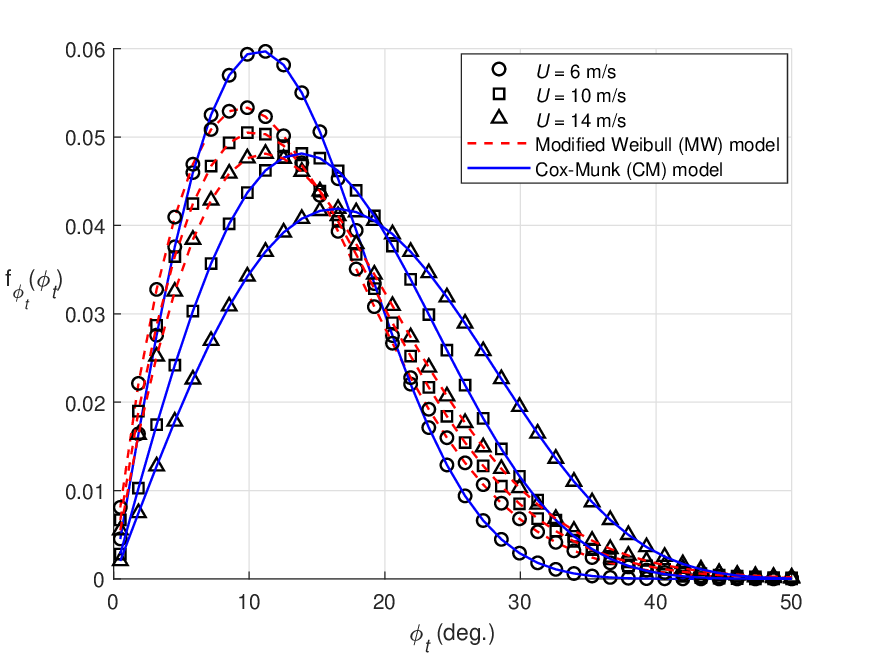}
    \caption{\small Normalized PDF of transmitter tilt angle $\phi_t$ for wind speeds, $U =$ 6, 10 and 14\,m/s under CM and \textcolor{black}{MW} sea-surface models.}
	\label{fig:CM_MW_windspeeds}
\end{figure}


For completeness, we have compared in Fig.~\ref{fig:CM_MW_windspeeds} the statistical behavior of the air-sea interface through the Tx tilt angle distribution $\phi_t$, using the CM and \textcolor{black}{MW} PDFs in (\ref{eq:phitpdf_CM}) and (\ref{eq:phitpdf_IE}), respectively.
Although the CM model exhibits slightly more variability (likely reflecting differences in measurement conditions), both models show similar qualitative trends across wind speeds, confirming their physical consistency. As wind speed increases from $6$ to $14$~m/s, angular variance increases, resulting in broader PDFs with lower peaks, indicative of rougher surfaces and steeper slopes. Quantitatively, the most probable tilt angle shifts from about $11^\circ$ to $17^\circ$ for CM and from $10^\circ$ to $11^\circ$ for \textcolor{black}{MW}, while heavier tails indicate a non-negligible probability of severe misalignment events. For example, under CM at $U=14$~m/s, significant probability mass persists beyond $30^\circ$, whereas at $U=6$~m/s the PDF rapidly diminishes beyond $30^\circ$. 

\section{Ergodic Capacity Analysis}\label{Sec-Metric}
To assess the performance of an underwater-to-air OWC link, we consider here the performance metric of ergodic channel capacity.
For an intensity modulation with direct detection (IM/DD) link, the instantaneous channel capacity (in bps/Hz) can be approximated as \cite{ijeh2022ergodic}:
\begin{equation}\label{Equ:InstCerg}
	\mathcal{C} \approx \frac{1}{2}\log_2\!\left(1+\frac{e}{2\pi}\gamma\right),
\end{equation}
where $e$ is the electron charge, and \( \gamma \) denotes the instantaneous electrical signal-to-noise ratio (SNR) at the Rx, given by $\gamma = \mu^2 h^2/(\alpha h + \beta)$,
where \( \mu = R_L {\cal R}_\text{e} P_{\mathrm{Tx}} \) with \( {\cal R}_\text{e} \) being the PD responsivity, \( R_L \) the TIA feedback resistor, and $P_{\mathrm{Tx}}$ the peak transmit optical power. Also, \( \alpha = 2\,e\,G\,F\,B_e\,R_\text{L}\,\mu \) accounts for signal-dependent noise contribution, dominated by shot noise, while \( \beta = R_\text{L}^2\,2\,e\,G\,F\,B_e\, (I_{\text{d}} + I_{\text{b}}) + 4\,K_\text{Z}\,T\,B_e\,R_\text{L}\) aggregates signal-independent noise sources, including dark current noise, background-induced shot noise, and thermal noise (see Table~\ref{table:sim. parameters} for parameters definitions).

The ergodic capacity is obtained by averaging the instantaneous capacity over all channel realizations \textcolor{black}{as \cite{ijeh2022ergodic}:} 
\begin{equation}\label{Equ:AvgCerg}
	\mathcal{C}_{\mathrm{erg}} \approx \frac{1}{2}
	\int_{0}^{\infty}
	\log_2\!\left(1+\frac{e}{2\pi}
	\frac{\mu^2 h^2}{(\alpha h + \beta)}\right)
	f_h(h)\, \mathrm{d}h,
\end{equation}
where \( f_h(h) \) denotes the PDF of the channel gain $h$ (see (\ref{Eq:h-complete})), which consists of a continuous fading term $h' \triangleq h_c \, h_{\phi_\text{t}} \, h_{\phi_\text{r}}$ and a discrete link interruption event $h_a$ caused by Rx FoV limitation. Accordingly, the PDF of \( h \) can be expressed as \cite{ijeh2022ergodic}:

{\small
\begin{equation}\label{Equ:fh}
		f_{h}{(h)} = f_{h'}{(h)}\underbrace{\int\limits_{0}^{\phi_\text{FoV}}f_{\phi_{r}}(\phi_{r})d{\phi_{r}}}_{\displaystyle P_{\mathrm{in}}}
		+ \underbrace{\bigg(1 - \int\limits_{0}^{\phi_\text{FoV}}f_{\phi_\text{r}}(\phi_\text{r})d{\phi_\text{r}}\bigg)}_{\displaystyle P_{\mathrm{out}}}\delta({h}),
\end{equation}}
where \( P_{\mathrm{in}} \), \( P_{\mathrm{out}} \) are the probabilities that the received beam lies inside or outside the Rx FoV, respectively. We have \cite{Ijeh-JOE-2021}: 
\begin{align}\label{fh2}
	f_{h'}{(h')}=\int f_{h'|h_{\phi_\text{r}}}{(h'|h_{\phi_\text{r}})}\ f_{h_{\phi_\text{r}}}({h_{\phi_\text{r}}})\ d{h_{\phi_\text{r}}},&&
\end{align} 
where the PDFs $f_{h_{\phi_\text{r}}}{(h_{\phi_\text{r}})}$ and $f_{h'|h_{\phi_r}}(h'|h_{\phi_r})$, given by Eqs. (44) and (45) in \cite{Ijeh-JOE-2021}, depend on PDFs $f_{\phi_t}(\cdot)$ and $f_{\phi_r}(\cdot)$ representing Tx misalignment (modeled by the proposed \textcolor{black}{MW} model) and Rx misalignment with \( \phi_r \sim \mathcal{N}(0,\sigma_{\phi_r}^2) \), respectively.



\begin{table}[t]
	\centering
	\caption{Simulation Parameters}
	\label{table:sim. parameters}
	\begin{tabular} {|    p{1.9265in}   |    p{1.05in}   |}
		\hline
		\textbf{Parameter} & \textbf{Value} \\   \hline \hline 
		LED wavelength   $\lambda$                   & $470\,\text{nm}$ \\  \hline
		LED peak transmit power $P_{\text{Tx}}$ &   $20\,\text{W}$ \\  \hline
		LED half-angle $\phi_{t_{1/2}}$ & $\approx 15^\circ$ at $m \approx 20$ \\  \hline
		Bit rate $R_\text{b}$ & $10\,\text{Mbps}$ \\  \hline
		Atmospheric transmittance (clear) $T_{a}$ & $\approx 0.98$ \\  \hline
		Optical filter transmittance $T_{f}$  & $1$ \\  \hline
		Optical filter bandwidth $B_o$  & $20\,\text{nm}$ \\  \hline
		Rx lens refractive index $n_{rf}$  & $1.5$ \\  \hline
		LPF bandwidth $B_e $ & $\approx R_\text{b}/2\ \textcolor{black}{\text{Hz}}$ \\  \hline
		SiPM active area  ${A}_\text{PD}$  &                                $9\,\text{mm}^2$\\ \hline
		SiPM, number of SPADs & 10998 \\ \hline
		SiPM gain $G$ &       $10^6$\\ \hline
		SiPM responsivity ${\cal R}_\text{e}$       &    $9 \times 10^{4}$ A/W \cite{Ijeh-JOE-2021} \\  \hline
		SiPM dark current  ${I}_\text{d}$  &                             $1.10\,\mu\text{A}$\\ 
		\hline
		SiPM excess noise factor $F$ &           $1.1$ \\  \hline
		TIA load resistance $R_\text{L}$ & $1\,\text{k}\Omega$ \\  \hline
		Diffuse attenuation coefficient in water $K_{w}$   & 0.08\,m$^{-1}$ \cite{Mobley-AP-1994} \\  \hline
		Diffuse attenuation coefficient in air $K_{a}$   & 0.19\,dB/km \cite{Ghassemlooy-CRC-2019}\\  \hline
		Total upwelling solar radiance $L_{t}(\lambda)$, for solar zenith angles $\theta>45^\circ$ and low sun-glint conditions ($0-0.01$) & $0.025\,\text{W}\text{m}^{-2}\text{nm}^{-1}\text{sr}^{-1}$ \cite{angara2024performance} 
		\\ \hline
		Solid angle of the
		optical filter $\Omega_{\phi_\text{FoVr}} = 2\,\pi\left(1 - \cos\left(\frac{\phi_\text{FoVr}}{2}\right)\right)$ & $\approx 0.2141\,\text{sr}$ at $\phi_\text{FoVr}$ of $30^\circ$
		\\ \hline
		Background current $I_b = {\cal R}_\text{e}\, L_{t}(\lambda)\,\Omega_{\phi_\text{FoVr}}\, T_a\, T_f\, B_o\, A_\text{PD}$ \cite{ijeh2022bit}  & $85\,\text{mA}$ \\
		\hline 
	\end{tabular}
\end{table}

\section{Numerical Results}\label{Sec-Results}
To evaluate the impact of air-sea surface statistics and system design parameters on the performance of an AUV-to-UAV OWC link, without loss of generality, we consider equal propagation distances in water and air, i.e., $Z_w = Z_a = Z/2$. Given the strong susceptibility of the ultrasensitive PD, i.e., the SiPM, to background noise, $L_{t}(\lambda)$ is selected for the conditions of minimal solar illumination \cite{angara2024performance}.
The remaining system and channel parameters used throughout the study are summarized in Table~\ref{table:sim. parameters}.
The ergodic capacity ${\cal C}_{\mathrm{erg}}$ is calculated analytically via numerical integration of (\ref{Equ:fh}) in (\ref{Equ:AvgCerg}), and validated through Monte Carlo simulations by averaging (\ref{Equ:InstCerg}) over $10^6$ independent channel realizations with randomly generated Tx and Rx tilt angles $(\phi_t,\phi_r)$. 


\subsection{Effect of Wind Speed and Link Range on Ergodic Capacity}

\begin{figure}
	\centering
	\includegraphics[width=\linewidth]{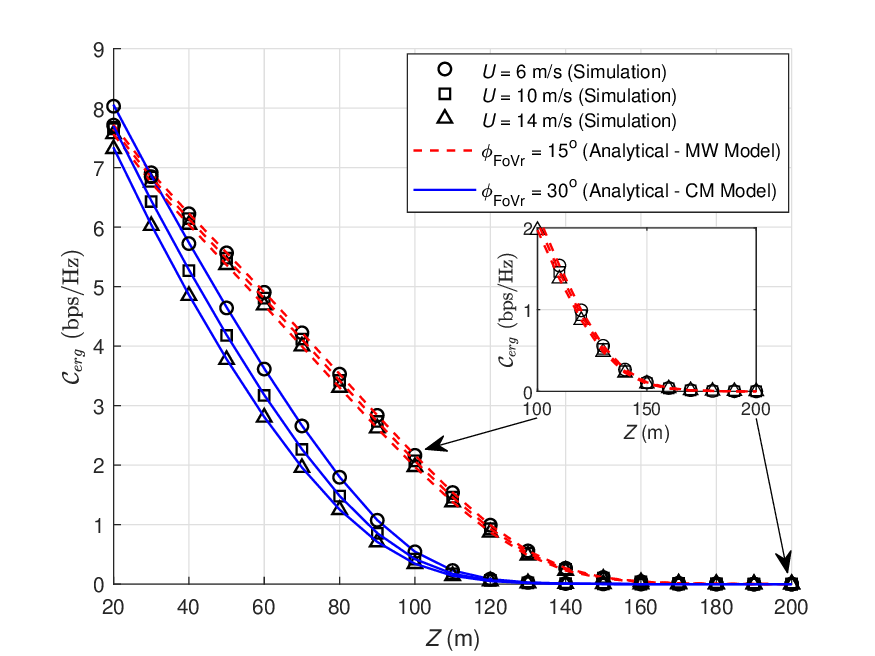}
	\caption{\small Effect of wind speed, Rx FoV, and operating range on ergodic capacity; $m=20$, $\sigma_{\phi_r}=10^\circ$, $L_{t}(\lambda) = 0.025\,\text{W}\text{m}^{-2}\text{nm}^{-1}\text{sr}^{-1}$.}
	\label{fig:CM_MW_Cerg_windspeeds}
\end{figure}

Figure~\ref{fig:CM_MW_Cerg_windspeeds} illustrates the impact of wind speed $U$, Rx FoV, and link range $Z$ on ergodic capacity while considering both IE and CM surface models. The excellent agreement between analytical and simulation results validates the proposed framework. For all configurations, ${\cal C}_{\mathrm{erg}}$ decreases monotonically with $Z$ due to increased path loss. At low to moderate ranges, lower $U$ yields higher capacity due to reduced surface-induced fading, while at larger $Z$, the link capacity becomes very small irrespective of $U$. Furthermore, the \textcolor{black}{MW} model evaluated with a narrow FoV $(\phi_{\mathrm{FoVr}}=15^\circ)$ outperforms the CM model with a wide FoV $(\phi_{\mathrm{FoVr}}=30^\circ)$, primarily due to reduced solar background noise, especially as the UAV moves farther away ($Z>20$\,m) from the sea surface. The selection of different FoV values for each model is intended for illustrative insight rather than comparative evaluation, as a direct comparison of the two models would not be meaningful. The subsequent analysis is performed based on the \textcolor{black}{MW} model only.

\subsection{Impact of Rx Angular Misalignment and Beam Size}

\begin{figure}[t]
	\centering
	\includegraphics[width=\linewidth]{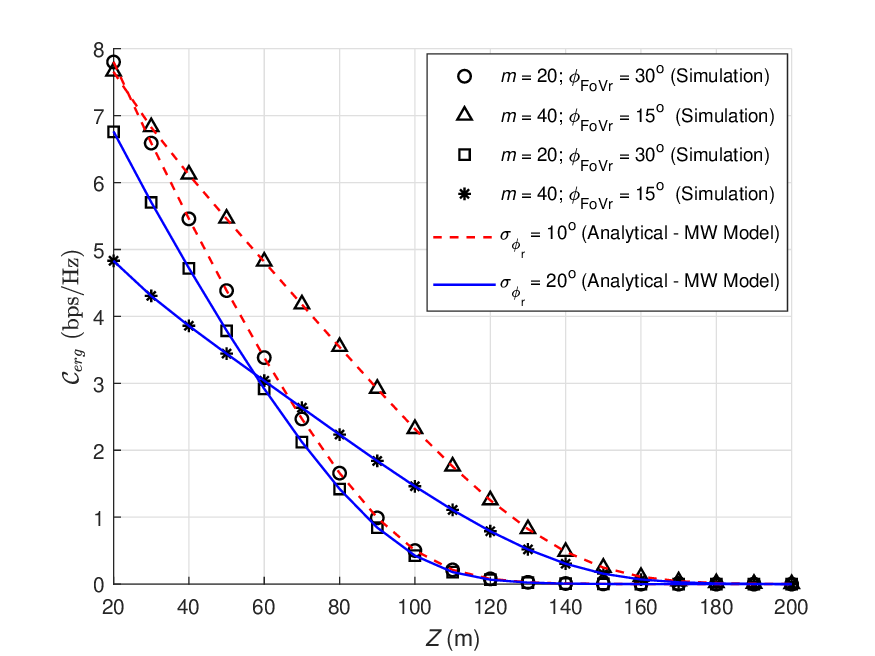}
	\caption{\small Effect of Rx angular misalignment $\sigma_{\phi_r}$, beam size $m$, FoV $\phi_{\mathrm{FoVr}}$, and range $Z$ on ${\cal C}_{\mathrm{erg}}$; $U=10$~m/s, $L_{t}(\lambda) = 0.025\,\text{W}\text{m}^{-2}\text{nm}^{-1}\text{sr}^{-1}$.}
	\label{fig:MW_sigmaphir}
\end{figure}

The influence of Rx angular misalignment is examined in Fig.~\ref{fig:MW_sigmaphir}. For small misalignment $(\sigma_{\phi_r}=10^\circ)$, a bow-shaped decay with link range is observed, where capacity is high at low ranges (e.g., $> 7$~bps/Hz at $Z=20$\,m) and decreases rapidly with $Z$, \textcolor{black}{suggesting suitability for applications in short-to-moderate link range.} In this regime, a smaller FoV consistently outperforms a larger FoV, regardless of beam divergence ($15^\circ$ and $11^\circ$ for $m = 20$ and $40$ respectively), due to reduced solar noise and a low probability of link interruption, \textcolor{black}{although practical LED beam spread and modulation limit achievable performance, in contrast to using a laser diode (LD).} In contrast, for stronger misalignment $(\sigma_{\phi_r}=20^\circ)$, an X-shaped trend emerges with three distinct phases: (i) for small link range ($<60$\,m), larger FoV provides higher capacity by relaxing angular alignment constraints; (ii) at moderate range ($50-60$\,m), an intersection point appears where both FoV configurations yield comparable performance, indicating an optimal operating range without adaptive parameter tuning; and (iii) at large distances ($>60$\,m), smaller FoV and narrower beam become advantageous, as solar noise and geometric loss dominate, \textcolor{black}{albeit with increased sensitivity to misalignment}.

\subsection{Effect of Solar Radiation and Receiver FoV}

\begin{figure}[t]
	\centering
	\includegraphics[width=\linewidth]{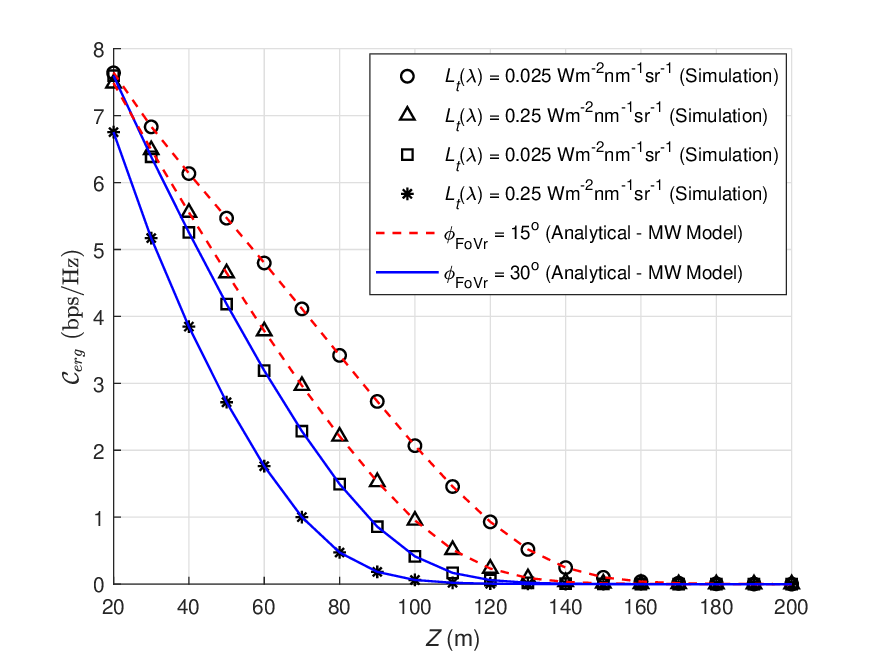}
	\caption{\small Effect of upwelling solar radiance $L_{t}(\lambda)$, FoV $\phi_{\mathrm{FoVr}}$, and range $Z$ on ${\cal C}_{\mathrm{erg}}$; $m=20$, $U=10$~m/s, $\sigma_{\phi_r}=10^\circ$.}
	\label{fig:CM_MW_Cerg_Esun}
\end{figure}

Finally, Fig.~\ref{fig:CM_MW_Cerg_Esun} highlights the impact of solar background radiation on the AUV-UAV OWC link. At low irradiance $L_{t}(\lambda) = 0.025\,\text{W}\text{m}^{-2}\text{nm}^{-1}\text{sr}^{-1}$ high link capacity is obtained across all link range, with improved performance achieved using a narrow FoV. For example, at $Z=60$\,m, with $\phi_{\mathrm{FoVr}}=15^\circ$ achieves approximately $5$\,bps/Hz, compared to $3$\,bps/Hz with $\phi_{\mathrm{FoVr}}=30^\circ$. Increasing the irradiance to $0.25\,\mathrm{Wm^{-2}nm^{-1}}\text{sr}^{-1}$ significantly degrades performance due to increased background noise, especially for larger FoVs. \textcolor{black}{This effect is further influenced by the SiPM receiver, whose high gain improves weak signal detection but also increases sensitivity to ambient noise. 
This introduces a trade-off between signal collection and noise mitigation, highlighting the importance of FoV selection and solar radiation in system design under moderate misalignment.}

\section{Conclusions}\label{Sec-Concl}
This work investigated a vertical underwater-to-air AUV-to-UAV optical link. We proposed a tractable formulation of the ECKV model validated by in-situ stereo image measurements, to account for sea surface-induced beam misalignments. Using this framework, the link ergodic capacity was evaluated under varying surface conditions and link ranges. To derive design insights, we analyzed the impact of link range, surface roughness, Rx misalignment, and FoV on capacity using analytical and numerical methods, highlighting the FoV trade-off between misalignment tolerance and solar noise collection.
\balance

\bibliographystyle{IEEEtran}
\bibliography{Ikenna_W2A}
\balance

\end{document}